\documentclass[doublecol]{epl2} 

\usepackage{graphicx}
\usepackage{bm,color,amsmath,amssymb}

\newcommand{\Tr}[1]{\underset{#1}{\mbox{Tr}}}
\newcommand{\extr}[1]{\underset{#1}{\mbox{extr}}}
\newcommand{\Eref}[1]{eq.~(\ref{#1})}
\newcommand{\Fref}[1]{Fig.~\ref{#1}}
\newcommand{\figwidth}{3.35in}

\title{Statistical Mechanics of Dictionary Learning}
\shorttitle{Statistical Mechanics of Dictionary Learning} 

\author{Ayaka Sakata\inst{1} \and Yoshiyuki Kabashima\inst{1}}
\shortauthor{Ayaka Sakata and Yoshiyuki Kabashima}

\institute{                    
  \inst{1} Department of Computational Intelligence and Systems Science, Tokyo Institute of Technology, Yokohama 226-8502, Japan.
}

\abstract{
Finding a basis matrix (dictionary) by which objective signals are represented sparsely 
is of major relevance in various scientific and technological fields. 
We consider a problem to learn a dictionary from a set of training signals. 
We employ techniques of statistical mechanics of disordered systems to evaluate the 
size of the training set necessary to typically succeed in the dictionary learning. 
The results indicate that the necessary size is much smaller than previously estimated, 
which theoretically supports and/or encourages the use of dictionary learning in practical situations. 
}

\begin{document}

\maketitle

\section{Introduction}

In various fields of science and technology,
such as earth observation, astronomy,
medicine, civil engineering, 
materials science, and in
compiling image databases \cite{Starck},
it has a major relevance to
recover original signals 
from deficient signals obtained by limited number of measurements. 
The Nyquist-Shannon sampling theorem \cite{Nyquist} 
provides the necessary and sufficient number of measurements 
for recovering arbitrary band-limited signals.
However, techniques based on this theorem sometimes do not match restrictions and/or demands of today's 
front-line applications \cite{front-line,Protein}, and much effort is still 
being made to find more efficient methodologies. 

The concept of sparse representations has recently drawn great attention in such research. 
Many real world signals such as natural images are represented {\em sparsely} 
in Fourier/wavelet domains; namely, many components vanish or are negligibly 
small in amplitude when the signals are represented by Fourier/wavelet bases. 
This empirical property is exploited in the 
signal recovery paradigm of compressed sensing (CS)
enabling the recovery of
sparse signals from much fewer measurements than 
those the sampling theorem estimates \cite{Donoho,Candes,Kabashima,Montanari,Sompolinsky,Krzakala}.

However, the effectiveness of
CS relies considerably on the assumption that a basis 
by which the objective signals look sparse is known in advance. 
Therefore, 
in applying CS to general 
signals of interest,
whose bases for sparse representation are
unknown, the primary task to accomplish 
is to identify 
an appropriate basis (dictionary) for the sparse representation from an available set of training signals. 
This is often termed {\em dictionary learning (DL)} \cite{Rubinstein,Elad,Gleichman}. 

Let us denote the training set of $M$-dimensional signals as an 
$M \times P$ matrix $\bm{Y}=\{Y_{\mu l}\}$,
where each column vector $\bm{Y}_l$
represents a sample signal and $P$ is the number of 
the samples.
In a simple scenario, DL is formulated as a problem to find a pair of an $M \times N$ matrix (dictionary) 
$\bm{D}\!=\!\{D_{\mu i}\}$ and an $N \times P$ sparse matrix $\bm{X}\!=\!\{X_{i l}\}$ such that 
$\bm{Y}\!=\!\bm{D} \bm{X}$ holds.
By DL, 
the characteristics/trends underlying $\{\bm{Y}_l\}$ 
are extracted into $\bm{D}$, and $\bm{Y}_l$ can be 
compactly represented as a superposition of 
a few dictionary columns, 
whose combination and strength are 
specified by the sparse matrix $\bm{X}$. 
DL suits not only efficient signal processing 
such as CS, but also extraction of non-trivial regularities from 
high-dimensional data.
For instance, DL has been successfully
applied to the facial image processing
for the efficient storage of 
large databases, where standard
algorithms fail \cite{Bryt-Elad, Elad}.
In this case, 
$\bm{Y}_l$ and $\bm{D}$ correspond to a 
facial image and a collection of patches 
of facial patterns learned by the $P$ samples, respectively. 
A variant of DL has also been employed in gene expression analysis
to estimate transcription factor activity $\bm{D}$ from 
gene expression data 
of a small size 
$\{\bm{Y}_l\}$ \cite{Gong2011}.

An important question of DL
is how large a sample size $P$ is necessary to 
uniquely identify an appropriate dictionary $\bm{D}$,
because the ambiguity of the dictionary is fatal 
in use for signal/data analysis after learning.
As the first answer to this question, 
an earlier study 
based on linear algebra showed that when the training set $\bm{Y}$ is generated by a pair of 
matrices $\bm{D}^0$ and $\bm{X}^0$ (planted solution) as $\bm{Y}=\bm{D}^0 \bm{X}^0$, 
one can perfectly learn these as a unique solution
except for the ambiguities of signs and permutations of matrix elements 
if $P > P_{\rm c}=(k+1) {}_N C_k$ and $k$ is sufficiently small, where $k$ is the number of 
non-zero elements in each column of $\bm{X}^0$ \cite{Aharon}. 
This result is significant as 
it is the first proof that guarantees the learnability with a finite size sample set for DL. 
However, the estimate of $P_{\rm c}$ is supposed to enable a considerable improvement; the authors 
of \cite{Aharon} speculated that $P_{\rm c}$ could be reduced substantially 
to $O(N^2)$ or even 
smaller, although providing a mathematical proof was technically difficult. 
The improvement of the estimation $P_{\rm c}$
is practically significant
because it will lead to considerable reduction of necessary cost for DL
in terms of both sample and computational complexities. 

In this Letter, we take an alternative approach to estimating $P_{\rm c}$. 
Specifically, we examine the typical behavior of DL using the replica method in the limit 
of $N,M,P \to \infty$. The main result of our analysis is that the planted solution is 
typically learnable by $O(N)$ training samples if negligible mean square errors per 
element are allowed and $M/N$ is sufficiently large. This theoretically supports and/or 
encourages the employment of DL in practical applications. 

\section{Problem setting}

We focus on the learning strategy
\begin{eqnarray}
&&\min_{\bm{D},\bm{X}}||\bm{Y}(=N^{-1/2}\bm{D}^0\bm{X}^0)-N^{-1/2} \bm{D}\bm{X}||^2 \cr 
 && \ \ \mbox{subj. to} \ ||\bm{D}||^2 = MN, \ ||\bm{X}||_0=NP \theta
\label{DL_optimization}
\end{eqnarray}
\cite{Aharon,Olshausen-Field,Engan,K-SVD,Rubinstein,Elad}, where 
$||\bm{A}||^2\!=\!\sum_{\mu l} A_{\mu l}^2$ for a matrix $\bm{A}\!=\!\{A_{\mu l} \}$, 
and $||\bm{A}||_0$ represents the number of non-zero elements of $\bm{A}$. 
The parameter $\theta \in [0,1]$ denotes the rate of non-zero elements assumed by 
the learner, and $N^{-1/2}$ is introduced for convenience in taking the large system limit. 

For simplicity, we assume that $\bm{D}^0$ and $\bm{X}^0$ of the planted solution 
are uniformly generated under the constraints of $||\bm{D}^0||^2=MN$, 
$||\bm{X}^0||_0=NP\rho$ and $||\bm{X}^0||^2=NP \rho$. We consider that 
the correct non-zero density $\rho$ can differ from $\theta$ for generality, 
but we assume $\rho \le \theta$; otherwise, the correct identification of 
$\bm{D}^0$ and $\bm{X}^0$ is trivially impossible. The main goal of our study 
is to evaluate the critical sample ratio $\gamma_{\rm c}=P_{\rm c}/N$ above which the planted solution can be learned typically. 

\section{Statistical mechanics approach}

Partition function 
\begin{align}
\nonumber
Z_\beta (\bm{D}^0,\bm{X}^0) &= \int d\bm{D}d\bm{X}\exp\left (-\frac{\beta}{2N}||\bm{D}\bm{X}-\bm{D}^0\bm{X}^0||^2
\right ) \\
&\times \delta(||\bm{D}||^2-NM)\delta(||\bm{X}||_0-NP\theta) 
\label{partition}
\end{align}
constitutes the basis of our approach since the minimized cost of \Eref{DL_optimization} can be identified with the zero temperature free energy 
${F}\!=\!-\lim_{\beta \to \infty} \beta^{-1} \ln Z(\bm{D}^0,\bm{X}^0;\beta)$. This statistically fluctuates depending on $\bm{D}^0$ and $\bm{X}^0$. 
However, as $N,M,P \to \infty$, 
one can expect that the {\em self-averaging} property is realized; i.e., the free energy density 
$N^{-2} {F}$ converges to the typical value $f\equiv N^{-2} [{F}]_0$ with probability unity, 
where $[\cdots ]_0$ stands for the average with respect to $\bm{D}^0$ and $\bm{X}^0$. 
Consequently, this property is also expected to hold for other relevant macroscopic 
variables of the solution of \Eref{DL_optimization}, $\bm{D}^*$ and $\bm{X}^*$. 
Therefore, assessing $f$ is the central issue in our analysis. 

\begin{figure}
\begin{center}
\includegraphics[width=1.6in]{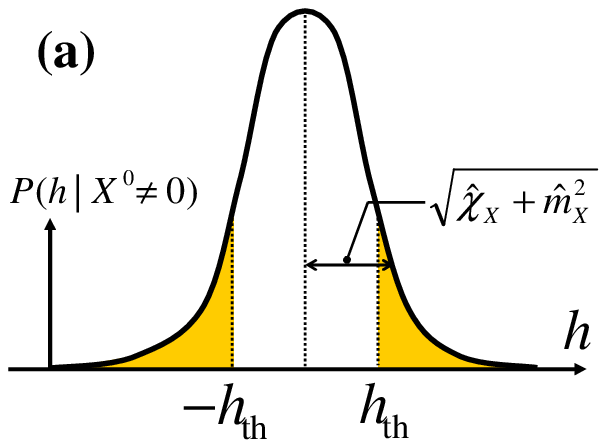}
\includegraphics[width=1.6in]{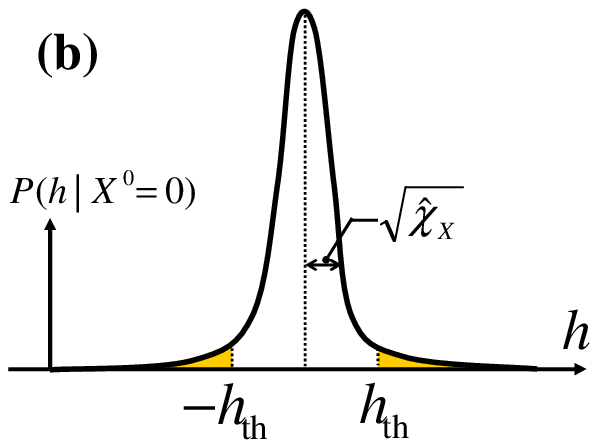}

\vspace{0.3cm}

\includegraphics[width=1.6in]{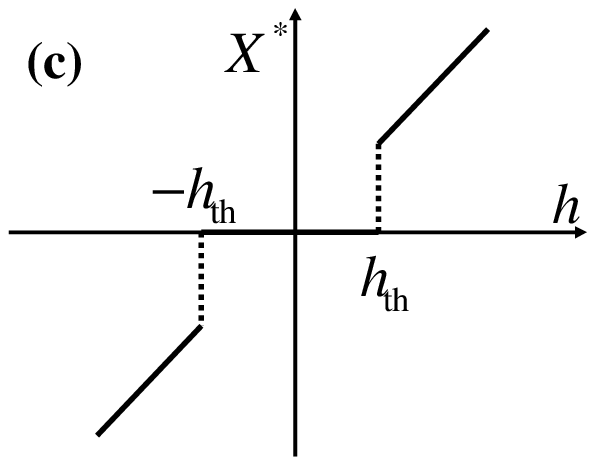}
\includegraphics[width=1.6in]{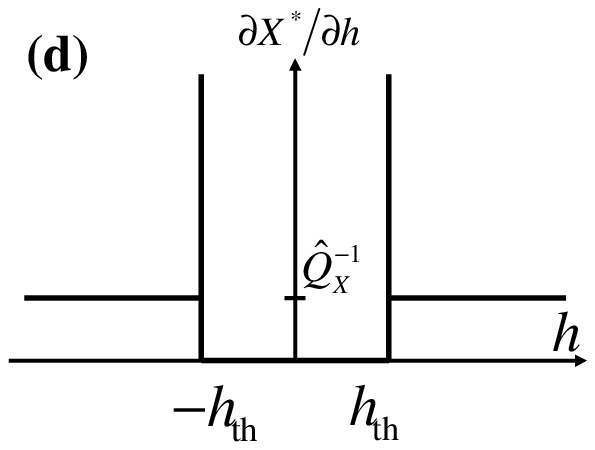}
\end{center}
\caption{(color online) (a) and (b) show distributions of local field $h$ (a) $P(h|X^0\neq 0)$ for $X^0\neq 0$ and (b) $P(h|X^0=0)$ for $X^0=0$. (c) and (d) show $X^*$ and $\partial X^*/\partial h$ as functions of $h$, respectively.}
\label{h_eff}
\end{figure}

This assessment can be carried out systematically using the replica method \cite{Dotzenko} 
in the limit of $N \to \infty$ while keeping $\alpha=M/N \sim O(1)$ and $\gamma=P/N\sim O(1)$. 
Under the replica symmetric (RS) ansatz, where the solution space of \Eref{DL_optimization} is assumed to 
be composed of at most a few pure states \cite{beyond}, the free energy density is given as 
\begin{align}
&f=\mathop{\rm extr}_{\Omega,\hat{\Omega}}\Big\{
\!-\!\alpha\Big(\frac{\hat{Q}_D\!-\!\hat{\chi}_D\chi_D}{2}\!-\!\hat{m}_Dm_D\!+\frac{\hat{\chi}_D\!+\!\hat{m}_D^2}{2\hat{Q}_D}\Big)\cr
&-\! \gamma\Big(\frac{\hat{Q}_XQ_X\!-\!\hat{\chi}_X\chi_X}{2} \!-\!\hat{m}_Xm_X\!+\!\lambda\theta\!-\!
\langle \langle\phi(h;\hat{Q}_X,\lambda) \rangle \rangle_{h}\Big)\cr
&+\frac{\alpha\gamma(Q_X-2m_Dm_X+\rho)}{2(1+Q_X\chi_D+\chi_X)}\Big\},
\label{free_energy}
\end{align}
where $\mathop{\rm extr}_{\Omega,\hat{\Omega}}\{{\cal G}(\Omega,\hat{\Omega})\}$ 
stands for the extremization of a function ${\cal G} (\Omega,\hat{\Omega})$ with respect to 
a set of macroscopic variables $\Omega\!\equiv \!\{\chi_D,m_D,Q_X,\chi_X,m_X\}$ and that of 
their conjugates $\hat{\Omega}\!\equiv \!\{\hat{Q}_D,\hat{\chi}_D,\hat{m}_D,\hat{Q}_X,\hat{\chi}_X,\hat{m}_X,\lambda\}$, and
\begin{align}
\phi(h;\hat{Q}_X, \lambda)=\min_X \lim_{\epsilon \to +0} \Big\{\frac{\hat{Q}_XX^2}{2}\!- \! hX\!+ \!\lambda |X|^\epsilon\Big\}.
\label{phi}
\end{align}
Notation $\langle \langle \cdots \rangle \rangle_h$ represents the average with respect to $h$ 
according to the distribution $P(h)=\rho P(h|X^0\neq 0)+(1-\rho)P(h|X^0=0)$, 
where $P(h|X^0\neq 0)$ and $P(h|X^0=0)$ are given by zero-mean Gaussian distributions 
with variances $\hat{\chi}_X+\hat{m}_X^2$ and $\hat{\chi}_X$, respectively (\Fref{h_eff}(a),(b)).
The details of the derivation of the free energy density 
are shown in Appendix.

\section{Physical implications} 

At the extremum of \Eref{free_energy}, the relationships 
\begin{align}
m_D&=\frac{1}{MN}[{\rm Tr}(\bm{D}^0)^{\rm T} \bm{D}^*]_0,\\
m_X&=\frac{1}{NP}[{\rm Tr}(\bm{X}^0)^{\rm T} \bm{X}^*]_0,\\
Q_X&=\frac{1}{NP}[{\rm Tr}(\bm{X}^*)^{\rm T} \bm{X}^*]_0=\frac{1}{NP}[||\bm{X}^*||^2]_0
\end{align}
hold, where {\rm T} denotes the matrix transpose. These provide the mean square errors (per element), which measure the performance of DL, as 
\begin{align}
\epsilon_D &\equiv \frac{1}{MN}[||\bm{D}^*-\bm{D}^0||^2]_0=2(1-m_D)\\
\epsilon_X &\equiv (NP)^{-1}[||\bm{X}^*-\bm{X}^0||^2]_0=\rho-2m_X+Q_X.
\end{align}
The variables $\chi_D$ and $\chi_X$ physically mean the sensitivity of the estimates 
$\bm{D}^*$ and $\bm{X}^*$ when the cost of \Eref{DL_optimization} is linearly perturbed.

Eq.(\ref{phi}) represents the effective single-body minimization problem concerning an 
element of $\bm{X}$ that is statistically equivalent to \Eref{DL_optimization} \cite{GuoVerdu}. 
Here, the randomness of $\bm{D}^0$ and $\bm{X}^0$ 
is effectively replaced by the random local field $h$. The first and second terms of 
$P(h)$ correspond to the cases where an element of $\bm{X}^0$ is given as $X^0\neq 0$ and 
$X^0=0$, respectively. Under a given $h$, the solution $X^*$ that minimizes the cost of \Eref{phi} 
is offered as $X^*=h/\hat{Q}_X$ for $|h| > h_{\rm th} \equiv (2 \hat{Q}_X \lambda)^{1/2}$ and $0$ 
otherwise (\Fref{h_eff} (c)). We refer to the cases of $|h| > h_{\rm th}$ and $|h| < h_{\rm th}$ 
as ``active'' and ``inactive,'' respectively. When $X^0\neq 0$, $h$ is generated from a Gaussian 
distribution ($P(h|X^0\neq 0)$) of zero-mean and variance $\hat{\chi}_X+\hat{m}_X^2$, and $X^*$ 
is more likely to be active than when $X^0=0$, for which $h$ is characterized by another zero-mean 
Gaussian ($P(h|X^0=0)$) of a smaller variance $\hat{\chi}_X$ (\Fref{h_eff} (a),(b)). Therefore, 
one can expect that the hard-thresholding scheme based on $h_{\rm th}$ represents proper assignment 
of zero/non-zero elements in $\bm{X}^*$ so as to accurately estimate $\bm{X}^0$ and $\bm{D}^0$ if $\hat{m}_X$ 
is sufficiently large.

A distinctive feature of $X^*$ is the divergence of the local susceptibility $\partial X^*/\partial h$ 
at ``border'' cases of $h=\pm h_{\rm th}$ (\Fref{h_eff} (d)). This affects the increase in the effective 
degree of freedom (ratio) as follows: 
$\theta_{\rm eff}=\theta + \langle \langle h_{\rm th}\delta(|h|-h_{\rm th}) \rangle \rangle_{h}$, 
whereas $h_{\rm th}$ is determined so as to satisfy $\theta=\int_{|h| > h_{\rm th}} dh P(h)$ 
indicating the sparsity condition $||\bm{X}||_{0}=NP\theta$. 
The excess $\langle \langle h_{\rm th} \delta(|h|-h_{\rm th})\rangle \rangle_{h}$ 
is supposed to represent a combinatorial complexity for classifying each element of 
$\bm{X}^*$ that corresponds to the border case $|h|=h_{\rm th}$ into the active case, 
$|h|>h_{\rm th}$ and $\bm{X}^*\neq0$, or the inactive case, $|h|<h_{\rm th}$ and $\bm{X}=0$. 
The divergence of $\partial X^*/\partial h|_{h=\pm h_{\rm th}}$ is also accompanied by the instability of 
the RS solution against perturbations that break the replica symmetry \cite{AT}. The influence of this instability is discussed later. 

\section{Actual solutions} 

We found two types of solutions; the first one is characterized by $m_D=1$ and $Q_X=m_X=\rho$, 
while the second is characterized by $m_X=0$ and $m_D=0$. The former case provides 
$\epsilon_D=\epsilon_X=0$ indicating the correct identification of $\bm{D}^0$ and $\bm{X}^0$, 
and hence we call it the {\em success} solution. The latter is referred to as the {\em failure} 
solution since $m_D=0$ and $m_X=0$ indicate the complete failure of information extraction of $\bm{D}^0$ and $\bm{X}^0$. 

\begin{figure}
\begin{center}
\includegraphics[width=\figwidth]{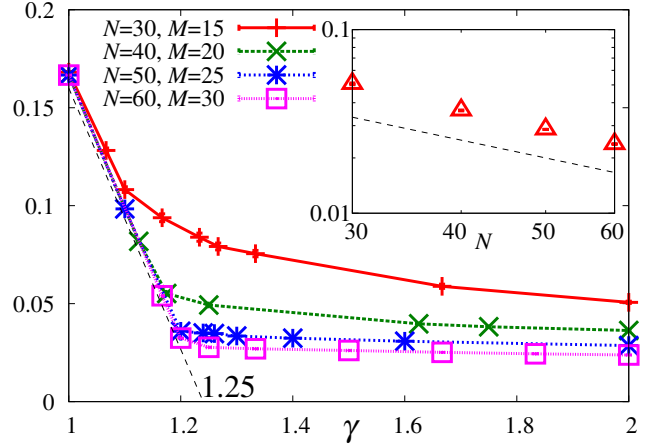}
\end{center}
\caption{(color online) $\gamma$-dependence of the ratio of marginal modes relative to $N^2$ at $\alpha=0.5$ and $\rho=\theta=0.1$. The behavior at $N\to\infty$ extrapolating from the results of finite $N$ is denoted by the dashed line. Inset: $N$-dependence of the ratio of marginal modes for $\gamma=2$. The dashed line stands for $N^{-1}$ as a guide. Each marker represents the average of 100 experiments.}
\label{zero-mode}
\end{figure}

{\em Success solution} ({\bf S}) exists when $\gamma>1$ and 
\begin{align}
\alpha > \theta_{\rm eff}^{\rm S}(\theta,\rho)= \theta+(1-\rho)\sqrt{\frac{2}{\pi}}u \exp\left(-\frac{u^2}{2}\right)
\end{align}
hold, where $u=H^{-1}( (\theta-\rho)/(2(1-\rho)))$ and $H^{-1}(x)$ is the inverse function of $H(x)=(2 \pi)^{-1/2}\int_x^\infty dt e^{-t^2/ 2}$. 
{\bf S} is further classified into two cases depending on $\gamma$. For $\gamma > \gamma_{\rm S}$, 
where 
\begin{align}
\gamma_{\rm S}(\alpha,\theta,\rho)=\frac{\alpha}{\alpha-\theta_{\rm eff}^{\rm S}}\label{gamma_S}, 
\end{align}
$\chi_D$ and $\chi_X$ are finite. On the other hand, for $1< \gamma < \gamma_{\rm S}$, $\chi_D$ and $\chi_X$ tend to infinity, keeping 
$R=\chi_D/\chi_X$ finite. 

To physically interpret this classification, let us take a variation around $\bm{Y}=N^{-1/2} \bm{D}^0\bm{X}^0$, which yields
\begin{align}
0=\delta(\bm{D}\bm{X})|_{\bm{D}^0,\bm{X}^0}=\bm{D}^0\delta\bm{X}+\delta\bm{D}\bm{X}^0. 
\label{eq_marginal}
\end{align}
If $\delta \bm{D}=0$ and $\delta \bm{X}=0$ are the unique solution of \Eref{eq_marginal}, the planted solution is locally stable. Otherwise, there are ``marginal'' modes along which the cost of \Eref{DL_optimization} does not increase locally, and the solution set forms a manifold. The number of constraints of \Eref{eq_marginal}, $MP$, coincides with that of the degree of freedom of $\delta\bm{D}$ and $\delta\bm{X}$, $MN+NP\theta_{\rm eff}^{\rm s}$, at $P=\gamma_{\rm S}N$. Thus, the classification below/above $\gamma_{\rm S}$ corresponds to the change in the number of marginal modes around the planted solution.

To confirm the validity of this interpretation, we numerically evaluated the number of marginal modes of \Eref{eq_marginal} in the case of $\alpha=1/2$ and $\theta=\rho=0.1$, which is shown in \Fref{zero-mode}. The assessment of $\gamma_{\rm s}$ when $\theta=\rho$ is conjectured to be exact since the effect of the border elements is negligible 
under this condition. \Fref{zero-mode} indicates that the number of marginal modes scales as $O(N^2)$ for $\gamma < \gamma_{\rm S}=1.25$, while it scales as $O(N)$, and the contribution of the marginal modes approaches zero, for $\gamma > \gamma_{\rm S}$ (inset). This result 
coincides with our theoretical assessment. At the same time, this implies that identifying the planted solution without any errors by \Eref{DL_optimization} is difficult as long as $\gamma \sim O(1)$, {but} the discrepancies per element caused by the marginal modes are negligibly small and could be allowed in many practical situations. 

In the case of $\gamma < 1$, for any $N \times P$ matrix $\bm{Z}$ of $||\bm{Z}||_0=NP\theta$, $\bm{X}^*=a \bm{Z}$ and $ \bm{D}^*= a^{-1} \bm{Y}(\bm{Z}\bm{Z}^{\rm T})^{-1}\bm{Z}^{\rm T}$ minimize the cost of (\ref{DL_optimization}) to zero, where $a$ is determined such that $||\bm{D}^*||^2=MN$. 
This implies that the set of solutions of \Eref{DL_optimization} spreads widely, and the weight of the planted solution is negligibly small in the state space. This may be why {\bf S} disappears for $\gamma < 1$. 

{\em Failure solution} ({\bf F}) exists for $\forall{\gamma}\ge0$. 
If 
\begin{align}
\alpha < \theta_{\rm eff}^{\rm F}(\theta)=\theta+ \sqrt{\frac{2}{\pi}}v\exp\left(-\frac{v^2}{2}\right)
\end{align}
where $v=H^{-1}(\theta\slash 2)$ holds, {\bf F} always offers $\chi_D,\chi_X \to \infty$ making the free energy $f$ vanish. For $\alpha > \theta_{\rm eff}^{\rm F}$, on the other hand, $\chi_D$ and $\chi_X$ become finite implying that a single solution of \Eref{DL_optimization} is locally stable for most directions and offers 
$f >0$, if $\gamma$ is greater than 
\begin{align}
 \gamma_{\rm F}(\alpha,\theta)=\frac{\alpha}{(\alpha^{1/2}-(\theta_{\rm 
 eff}^{\rm F})^{1/2})^2}.
\end{align}
The inequality $\theta_{\rm eff}^{\rm F} \ge \theta_{\rm eff}^{\rm S}$ always holds because the influence of the border elements for {\bf F} is stronger than that for {\bf S}, which leads to $\gamma_{\rm S}\leq\gamma_{\rm F}$.

\begin{figure}[t]
\begin{center}
\includegraphics[width=\figwidth]{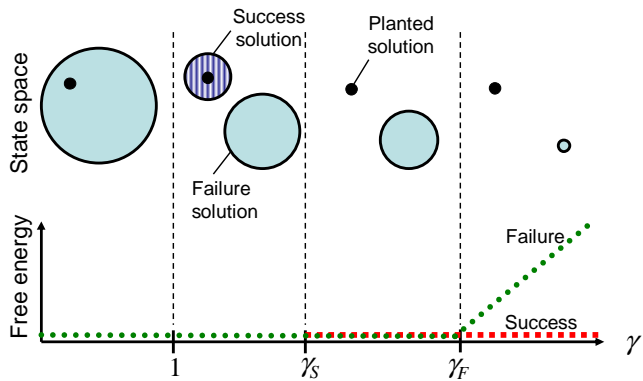}
\end{center}
\caption{(color online) Schematic pictures of $\gamma$-dependence of the phase space and free energy under RS assumption.}
\label{phase}
\end{figure}

\Fref{phase} illustrates changes in state space that occur for sufficiently large $\alpha$ under the RS assumption. For $\gamma < 1$, {\bf F} is a unique solution. As $\gamma$ increases, {\bf S} appears at $\gamma=1$, and the number of marginal modes changes from $O(N^2)$ to $O(N)$ at $\gamma =\gamma_{\rm S}$. This implies that when 
negligibly small linear perturbations are added to the cost of \Eref{DL_optimization}, the limits $\lim_{N\to\infty}\epsilon_D\sim 0$ and $\lim_{N\to\infty}\epsilon_X\sim0$ still hold for ${\bf S}$ of $\gamma > \gamma_{\rm S}$ while 
they can be boosted to $O(1)$ for {\bf S} of $\gamma < \gamma_{\rm S}$. For $\gamma < (\gamma_{\rm S} \le )\gamma_{\rm F}$, {\bf S} and {\bf F} are degenerated providing $f=0$. However, at $\gamma=\gamma_{\rm F}$, {\bf S} becomes thermodynamically dominant by keeping $f=0$, while {\bf F} begins to have positive $f$. This means that the planted solution is typically learnable by $P > P_{\rm c}=N\gamma_{\rm F} \sim O(N)$ training samples if negligible mean square errors per element are allowed. 

\begin{figure}
\begin{center}
\includegraphics[width=\figwidth]{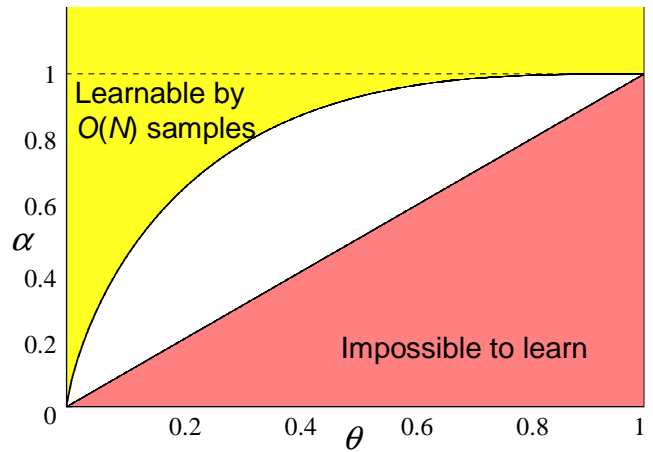}
\end{center}
\caption{(color online) Phase diagram on $\alpha-\theta$ plane.}
\label{alpha-theta}
\end{figure}

\Fref{alpha-theta} plots the phase diagram on an $\alpha-\theta$ plane. 
The region above $\alpha=\theta_{\rm eff}^{\rm F}(\theta)$ (curve) 
represents the condition under which the planted solution is typically 
learnable by $O(N)$ training samples. DL is impossible in the region below 
$\alpha =\theta$ (straight line) because $\bm{X}^0$ cannot be correctly 
recovered even if $\bm{D}^0$ is known \cite{Kabashima}. 
How the sample complexity scales with respect to
$N$ in the region $\theta < \alpha < \theta_{\rm eff}^{\rm F}(\theta)$ 
is beyond the scope of this Letter, but an interesting question nonetheless. 

\section{Summary and discussion}
In summary, we have assessed the size of training samples required for 
correctly learning a planted solution in DL using the replica method. 
Our analysis indicated that $O(N)$ samples, 
which are much fewer than estimated in an earlier study \cite{Aharon}, 
are sufficient for learning a planted dictionary with allowance for 
negligible square discrepancies per element when the number of non-zero 
signals is sufficiently small compared to that of measurements. 

It was shown 
that the identification of dictionary can be 
characterized 
as a phase transition with respect to the 
number of training samples.
Our RS analysis probably does not describe the exact behavior of DL since the 
RS solutions are unstable against the replica symmetry breaking (RSB) disturbances. 
However, we still speculate that the RS estimate of $\gamma_{\rm F}$ serves as an 
upper bound of the correct critical ratio $\gamma_{\rm c}$. This is because the 
free energy value of {\bf F} assessed under the RSB ansatz should be greater 
than or equal to that of the RS solution due to the positivity constraint of 
the entropy of pure states (complexity) \cite{Mezard-Montanari}, whereas that 
of {\bf S} is kept to vanish, which always yields a smaller estimate of $\gamma_{\rm F}$. 

Promising future research includes an extension of the current framework 
to more general situations such as noisy cases as well as refinement of 
the estimates of the critical ratios $\gamma_{\rm S}$ and $\gamma_{\rm F}$ 
taking RSB into account \cite{SK}. 

\acknowledgments
This work was partially supported by 
a Grant-in-Aid for JSPS Fellow (No. 23--4665) (AS)
and KAKENHI No. 22300003 (YK).

\appendix
\section{Appendix: Derivation of \Eref{free_energy}}
In general, the configurational average 
of the free energy density could be evaluated on the 
basis of the following formula:
\begin{align}
f_\beta=-\frac{1}{\beta}
\lim_{N\to\infty}\frac{1}{N^2}\lim_{n\to 0}
\frac{\partial }{\partial n}\ln [Z_\beta^n(\bm{D}^0,\bm{X}^0)]_{0}.
\label{replica_free_energy}
\end{align}
Unfortunately, assessing $[Z_\beta^n(\bm{D}^0,\bm{X}^0)]_{0}$
for $n \in \mathbb{R}$ in the mathematically rigorous manner
is technically difficult, 
%
and this fact
prohibits us from  
utilizing \Eref{replica_free_energy} in practice. 
In the replica method, this difficulty is resolved by
evaluating $N^{-2}\ln [Z_\beta^n(\bm{D}^0,\bm{X}^0)]_{0}$ for $n \in \mathbb{N}$ 
as an analytic function of $n$ first 
in the limit of $N \to \infty$, and 
taking the $n \to 0$ limit afterward 
with use of the obtained analytic function for $n \in \mathbb{R}$ as well.

More precisely, 
we evaluate $ [Z_\beta^n(\bm{D}^0,\bm{X}^0)]_{0}$ by averaging 
the right hand side of an identity 
\begin{align}
&Z_\beta^n(\bm{D}^0,\bm{X}^0)=\cr
&\int \prod_{a=1}^n
\left \{d\bm{D}^a \bm{X}^a
\delta(||\bm{D}^a||^2-NM)
\delta(||\bm{X}^a||_0-NP\theta) \right \} \cr
&\hspace*{1cm} 
\times
\exp\left (-\frac{\beta}{2N} 
\sum_{a=1}^n ||\bm{D}^a\bm{X}^a-\bm{D}^0 \bm{X}^0||^2 \right ), 
\label{power_Z}
\end{align} 
which is valid for only $n \in \mathbb{N}$,
over the distributions of 
the planted solutions $\bm{D}^0$ and $\bm{X}^0$ 
that are given by
\begin{eqnarray}
P_{D^0}(\bm{D}^0)=\frac{1}{{\cal N}_D}\delta({|}|\bm{D}^0|{|}^2-NM)
\end{eqnarray}
and
\begin{eqnarray}
P_{X^0}(\bm{X}^0)=\prod_{i,l}\left\{(1-\rho)\delta(X_{i,l})+\frac{\rho}{\sqrt{2\pi}}\exp\left(-\frac{X_{il}^2}{2}\right)\right\},
\end{eqnarray}
respectively, where ${\cal N}_D$ is the normalization constant. 
In performing the integrals
{of $2(n+1)$ variables $(\bm{D}^0, \{\bm{D}^a\})$ and $(\bm{X}^0, \{\bm{X}^a\})$} 
that come out in this evaluation, we insert trivial identities
{with respect to all combinations of replicas $(a,b=0,1,2,\ldots,n)$}, 
\begin{eqnarray}
1=NM \int d q_{D}^{ab} 
\delta(\Tr{} (\bm{D}^a)^{\rm T}\bm{D}^b-NMq_{D}^{ab}),  
\end{eqnarray}
{and}
\begin{eqnarray}
1=NP \int d q_{X}^{ab} 
\delta(\Tr{}(\bm{X}^a)^{{\rm T}}\bm{X}^b-NPq_{X}^{ab})  
\end{eqnarray}
to the integrand.
Let us denote ${\cal Q}_D \equiv (q_D^{ab})$ and 
${\cal Q}_X \equiv (q_X^{ab})$, and introduce 
two joint distributions
\begin{align}
\nonumber
P_D&(\{\bm{D}^a\};{\cal Q}_D)=\frac{P_{D^0}(\bm{D}^0)}{V_D({\cal Q}_D)}\\
&\times\prod_{a=1}^n\delta(||\bm{D}^a||^2-NM)\prod_{a<b}\delta(\Tr{}(\bm{D}^a)^{\rm T}\bm{D}^b-NMq_D^{ab}),\label{app:P_D}\\
\nonumber
P_X&(\{\bm{X}^a\};{\cal Q}_X)=\frac{P_{X^0}(\bm{X}^0)\delta(
||\bm{X}^0||_0
-NP\rho)}{V_X({\cal Q}_X)}\\
&\times\prod_{a=1}^n\delta(
||\bm{X}^a||_0
-NP\theta)\prod_{a\leq b}\delta(\Tr{}(\bm{X}^a)^{{\rm T}}\bm{X}^b-NPq_X^{ab}),\label{app:P_X}
\end{align}
where $V_D({\cal Q}_D)$ and $V_X({\cal Q}_X)$ are
the normalization constants.
The above-mentioned computation provides the following expression:
\begin{align}
\nonumber
[Z_\beta^n&(\bm{D}^0,\bm{X}^0)]_{0}\\
\nonumber
&=\int d(NM{\cal Q}_D)d(NP{\cal Q}_X)V_D({\cal Q}_D)V_X({\cal Q}_X) \cr
&\times\left [\prod_{a=1}^n\exp\left(-\frac{\beta}{2}\sum_{\mu,l}{t_{\mu 
 l}^a}^2\right) \right]_{{\cal Q}_X,{\cal Q}_D}, 
\end{align}
where $t_{\mu l}^a={N}^{-1\slash 2}\sum_{i=1}^N(D_{\mu i}^a X_{i 
l}^a-D_{\mu i}^0X_{i l}^0)$. Notation  
$\left [ \cdots \right ]_{{\cal Q}_D,{\cal Q}_X}$ represents  
the average with respect to {$\{\bm{D}^a\}$ and $\{\bm{X}^a\}$
within the state space specified by ${\cal Q}_D$ and ${\cal Q}_X$,
whose distributions are given by eqs. (\ref{app:P_D}) and (\ref{app:P_X}). }
Distributions (\ref{app:P_D}) and (\ref{app:P_X}) are 
independent of each other, and provide each entry of $\{\bm{D}^a\}$ and $\{\bm{X}^a\}$ with 
zero mean and a finite variance.  
This allows us to utilize the central limit theorem indicating that we can handle $t_{\mu l}^a$ as multivariate 
Gaussian random variables that follow  
\begin{align}
P_t(\{t^{a}_{\mu l}\})\!=\!\prod_{\mu l}\!\frac{1}{\sqrt{(2\pi)^{n}\det{\cal 
 T}}}\!\exp\!\left(\!-\frac{1}{2}\sum_{a,b}t_{\mu l}^a({\cal 
 T}^{-1})^{ab}t_{\mu l}^b\!\right),
\end{align}
where ${\cal T}$ stands for an $n\times n$ matrix 
whose entries are given as ${\cal T}^{ab}=q_D^{ab}q_X^{{a}b}-{(q_D^{a0}q_X^{a0}+q_D^{b0}q_X^{b0})}+\rho$. 
Utilizing this and evaluating integrals of ${\cal Q}_X$ and ${\cal Q}_D$ by 
means of the saddle point method lead to an expression
\begin{align}
\nonumber
\lim_{N\to\infty}&\frac{1}{N^2}[Z^n_\beta(\bm{D}^0,\bm{X}^0)]_0=\extr{}
\Big[-\frac{\alpha\gamma}{2}\ln\det({\cal I}_n+\beta{\cal T})\\
\nonumber
&+\gamma\Big\{\frac{{\rm Tr}\hat{\cal Q}_X{\cal 
 Q}_X}{2}+\ln
 \left (\int\{\prod_{a=0}^ndX^\alpha \}P_{X^0}(X^0)e^{-\Xi} \right )\Big\}\\
&+ \alpha \Big\{\frac{{\rm Tr}\hat{\cal Q}_D{\cal Q}_D}{2}
-\frac{1}{2}\ln \det \hat{\cal Q}_D \Big\}\cr
&{+}n\lambda \theta +\frac{n\alpha}{2}\ln(2\pi)\Big]. 
\label{finite_n_evaluation}
\end{align}
Here, ${\cal I}_n$ represents the $n\times n$ identity matrix, 
auxiliary variables
$\hat{\cal Q}_D\equiv (\hat{q}_D^{ab})$ and $\hat{\cal Q}_X\equiv (\hat{q}_X^{ab})$ 
are introduced in evaluating $V_D(\cal{Q}_D)$ and $V_X(\cal{Q}_X)$ with use of the saddle point method, 
and $\Xi \equiv \frac{1}{2}\sum_{a,b=0}^n\hat{q}_X^{ab}X^aX^b+\lambda\sum_{a=1}^n\lim_{\epsilon \to +0} |X^a|^\epsilon$. 
Extremization should be taken with respect to $\lambda$ and four kinds of macroscopic variables
${\cal Q}_D$, ${\cal Q}_X$, $\hat{\cal Q}_D$, and $\hat{\cal Q}_X$. 

Exactly evaluating \Eref{finite_n_evaluation} should provide the correct leading order 
estimate of 
$N^{-2} \ln [Z^n_\beta(\bm{D}^0,\bm{X}^0)]_0$ for each of $n \in \mathbb{N}$. 
However, we here restrict the candidate of the dominant saddle point 
to that of the replica symmetric form as
\begin{align}
q_D^{ab}&=\left\{ \begin{array}{ll}
1, & a=b \\
q_D, & a\neq b, ~(a,b\neq 0)\\
m_D, & a= 0, b\neq 0 \\
\end{array} \right.\\
q_X^{ab}&=\left\{ \begin{array}{ll}
Q_X, & a=b \\
q_X, & a\neq b, ~(a,b\neq 0)\\
m_X, & a= 0, b\neq 0 \\
\end{array} \right.\\
\hat q_D^{ab}&=\left\{ \begin{array}{ll}
\hat Q_D, & a=b \\
-\hat q_D, & a\neq b, ~(a,b\neq 0)\\
-\hat m_D, & a= 0, b\neq 0 \\
\end{array} \right.\\
\hat q_X^{ab}&=\left\{ \begin{array}{ll}
\hat Q_X, & a=b \\
-\hat q_X, & a\neq b, ~(a,b\neq 0)\\
-\hat m_X, & a= 0, b\neq 0 \\
\end{array} \right.
\end{align}
so as to obtain an analytic expression with respect to $n$. 
This yields
\begin{align}
&\ln \det({\cal I}_n+\beta{\cal T})
=n \ln (1+\beta (Q_X-q_Dq_X))\cr
& \hspace*{1cm}+\ln \Big(
1+n \frac{\beta(q_Dq_X-2m_Dm_X+\rho)}{1+\beta(Q_X-q_Dq_X)} \Big), 
\label{app:RS1}
\end{align}
\begin{align}
&\frac{{\rm Tr}\hat{\cal Q}_X{\cal Q}_X}{2}+
\ln \left (\int\prod_{a=0}^n dX^a P_{X^0}(X^0)e^{-\Xi} \right ) \cr
&=\frac{n}{2}\hat{Q}_X Q_X-n\hat{m}_X m_X-\frac{n(n-1)}{2}\hat{q}_Xq_X \cr
&+\ln 
\langle \langle 
\Big (
\int dX e^{-\xi} \Big)^n \rangle \rangle_h,
\label{app:RS2}
\end{align}
{and}
\begin{align}
&\frac{{\rm Tr}\hat{\cal Q}_D{\cal Q}_D}{2}-
\frac{1}{2}\ln \det \hat{\cal Q}_D \cr
&=\frac{n}{2}\hat{Q}_D-n \hat{m}_D m_D -\frac{n(n-1)}{2} \hat{q}_Dq_D \cr
&-\frac{n}{2}\ln (\hat{Q}_D+\hat{q}_D)-\frac{1}{2}
\left (1-n\frac{\hat{q}_D+\hat{m}_D^2}{\hat{Q}_D+\hat{q}_D} \right ), 
\label{app:RS3}
\end{align}
where $\xi\equiv (\hat{Q}_X+\hat{q}_X)X^2/2-hX+\lambda 
\sum_{\epsilon \to +0}|X|^\epsilon$. 
Further, the following 
replacement of variables is convenient in handling our computation
in the limit of $\beta \to \infty$:    
$\hat{Q}_D+\hat{q}_D\to\beta\hat{Q}_D$, $\hat{q}_D\to\beta^2\hat{\chi}_D$,
$1-q_D\to\chi_D\slash\beta$, $\hat{Q}_X+\hat{q}_X\to\beta\hat{Q}_X$, $\hat{q}_X\to\beta^2\hat{\chi}_X$,
$Q_X-q_X\to\chi_X\slash\beta$, and $\lambda\to\beta\lambda$. 
In $\beta\to\infty$,
integral with respect to $X$ in \Eref{app:RS2}, $\int dX e^{-\xi}$, 
is replaced to $e^{-\beta \phi(h;\hat{Q}_X,\lambda)}$
by applying the saddle point method.
Inserting eqs. (\ref{app:RS1})--(\ref{app:RS3})
and the rescaled variables 
into \Eref{finite_n_evaluation}
offers the expression of the 
zero temperature free energy density (\ref{free_energy}). 

\end{document}